\documentclass[reprint,prx,twocolumn,groupedaddress,superscriptaddress]{revtex4-1}
\usepackage{graphicx,color}
\usepackage{verbatim}
\usepackage{amssymb}   
\usepackage{amsmath}
\usepackage{amsfonts}
\usepackage{mathdots}
\usepackage{hyperref}
\usepackage{epsfig}
\usepackage{hyperref}

\definecolor{myblue}{RGB}{46, 48,146}
\hypersetup{colorlinks=true,linkcolor=myblue,citecolor=myblue,urlcolor=myblue, linktocpage}
\usepackage{braket}
\renewcommand\bra[1]{{\langle{#1}|}}
\makeatletter
\renewcommand\ket[1]{%
	\@ifnextchar\bra{\k@t{#1}\!}{\k@t{#1}}%
}
\newcommand\k@t[1]{{|{#1}\rangle}}
\makeatother

\usepackage{bm}
\begin{document}
	\title{Creating Localized Majorana Zero Modes in Quantum Anomalous Hall Insulator/Superconductor Heterostructures with a 
		Scissor}
	\author{Ying-Ming Xie}
	\thanks{These authors contributed equally to this work}
	\affiliation{Department of Physics, Hong Kong University of Science and Technology, Clear Water Bay, Hong Kong, China}
	\author{Xue-Jian Gao}
	\thanks{These authors contributed equally to this work}
	\affiliation{Department of Physics, Hong Kong University of Science and Technology, Clear Water Bay, Hong Kong, China}
	\author{Tai-Kai Ng}
	\affiliation{Department of Physics, Hong Kong University of Science and Technology, Clear Water Bay, Hong Kong, China}
	\author{K. T. Law} 
	\thanks{Corresponding author.\\phlaw@ust.hk}
	\affiliation{Department of Physics, Hong Kong University of Science and Technology, Clear Water Bay, Hong Kong, China}
	\date{\today}
	\begin{abstract}
	In this work, we demonstrate that making a cut (a narrow vacuum regime) in the bulk of a quantum anomalous Hall insulator (QAHI) creates a topologically protected single helical channel with counter-propagating  electron modes, and inducing superconductivity on the helical channel through proximity effect will create Majorana zero energy modes (MZMs) at the ends of the cut. In this  geometry, there is no need for the proximity gap to overcome the bulk insulating gap of the QAHI to create MZMs as  in the two-dimensional QAHI/superconductor (QAHI/SC) heterostructures. Therefore, the topological regime with MZMs is greatly enlarged. Furthermore, due to the presence of a single helical channel, the generation of low energy in-gap bound states caused by multiple conducting channels is avoided such that the MZMs can be well separated from other in-gap excitations in energy. This simple but practical approach allows the creation of a large number of MZMs in devices with complicated geometry such as hexons for measurement-based topological quantum computation. We further demonstrate how braiding of MZMs can be performed by controlling the coupling strength between the counter-propagating electron modes. 	\end{abstract}
	\pacs{}
	\maketitle
	
\emph{Introduction.}---The search for Majorana zero-energy modes (MZMs) has been an important subject in condensed matter systems due to their exotic non-Abelian statistics and  potential applications in fault-tolerant topological quantum computation \cite{Ivanov,Nayak2008,Kitaev1,Alicea2011,Alicea_2016}. The MZMs are first proposed to exist in the vortex cores of $p+ip$ superconductor \cite{Read2000} and at the ends of 1D $p$-wave superconductors \cite{Kitaev2001}. Due to the lack of intrinsic $p$-wave superconductors in nature, intense efforts have been made on engineering topological superconductors that support MZMs using superconductors with conventional pairing. Notable examples of these systems include the heterostructures of superconductors and topological insulators \cite{Fu2008}, semiconductor nanowire \cite{Oreg2010,Jaysau2010,Roman2010,Roman2011}, ferromagnetic chain \cite{Choy2011,DanielLoss2013}, metallic surface states \cite{Potter2012,Yingming_gold} and quantum anomalous Hall insulators (QAHIs) \cite{Chuizhen,MacDonald2018}, \textit{etc.}.  After nearly a decade of intense study, tremendous progress has been made and signatures of MZMs have been observed in several experimental setups \cite{Mourik2012,Rokhinson2012,Das2012,Deng2012,Yazdani2014,Albrecht2016,Zhang2018,  Berthold, Fornieri2019,Ren2019, Jia2015,Jia2016, Dinghong1,Manna,Vaitiekenas}.  However, to realize practical topological quantum computation, a few important issues have to be overcome: (i) A platform with a large topological regime where MZMs can be easily realized is needed. (ii) The MZMs should be separated in energy from other quasiparticle excitations. (iii) A large number of MZMs can be created and non-Abelian operations can be performed. These existing platforms suffer from the very narrow topological parameter regimes or they are difficult to realize MZMs on a larger scale for the braiding operations needed in quantum computation. However, platforms that can satisfy all of the above conditions are still lacking. 

As pointed out by Kitaev \cite{Kitaev1}, the condition of realizing MZMs at the ends of a (quasi-)1D superconductor is indeed rather simple which is to induce superconducting pairing on an odd number of conducting channels in a metallic wire. However, in usual nanowire/s-wave superconductor heterostructures, for example, the Zeeman field needed to create an odd number of conducting channels at the Fermi energy can also polarize the spins of electrons and suppress superconducting pairing on the nanowires. Therefore, the strength of the applied magnetic field is limited and the chemical potential of the nanowire has to be fine-tuned precisely in the order of meV to create MZMs \cite{Mourik2012}. Previously, the authors proposed using Ising superconductors to replace conventional s-wave superconductors such that the equal spin Cooper pairs generated by the Ising spin-orbit coupling can induce pairing on fully spin-polarized nanowires \cite{law_2020}. This scheme enlarges the topological regime by one order of magnitude and at the same time maintaining a nearly field-independent pairing gap on the wire. However, this scheme has yet to be realized experimentally.

Recently, the discovery of QAHIs allows the realization of an odd number of conducting channels without applying external magnetic fields \cite{Chang,Checkelsky,Xufeng,Bestwick,Feng,Chang2015,Kandala2015,Qi,Qinglin,Shen}. It was proposed that inducing superconductivity on a QAHI could result in a topological superconductor with chiral Majorana modes \cite{Qi}. However, the superconducting gap must overcome the insulating gap of the QAHI in order to create the chiral Majorana modes which is very difficult to achieve. A recent experiment claimed that such chiral Majorana modes have been observed through transport experiments but the results are still under debate \cite{Wangjing2015,Qinglin,Shen,Jaysau2,xiaogang,Kayyalha}. 

\begin{figure}
	\centering
	\includegraphics[width=1\linewidth]{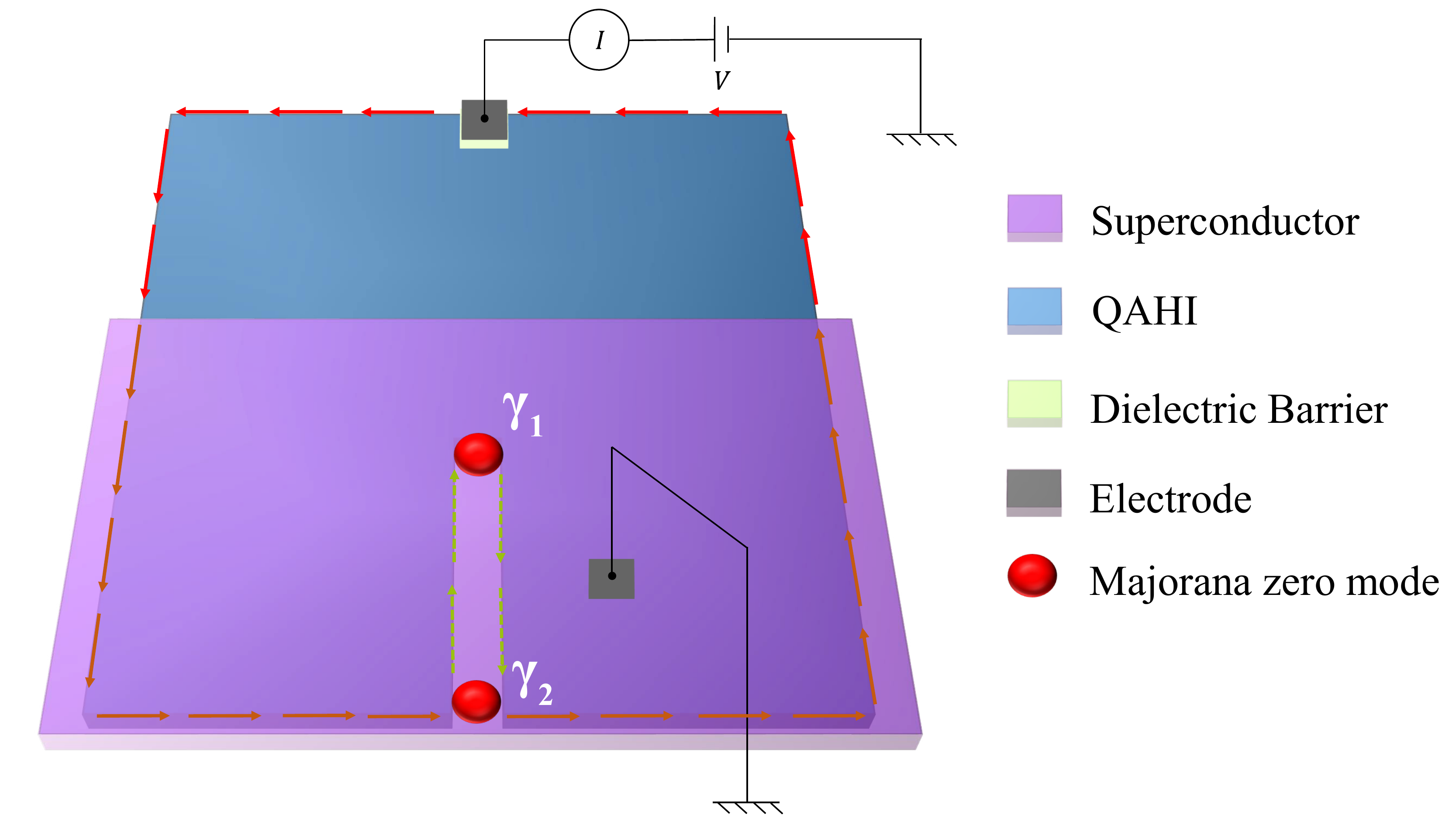}
	\caption{Schematically plot of the  QAHI/SC  heterostructure, where a wire-shaped region is removed from the QAHI. The superconductivity is induced by covering a superconductor on the top of the cut region. The red arrows at the edge of QAHI represent the chiral edge states of QAHI. The positions of MZMs $\gamma_1$ and $\gamma_2$ are denoted by red dots.  It is important to note that the MZM at the end of the cut is localized while the MZM at the beginning of the cut is delocalized due to the extended gapless chiral edge modes.}
	\label{fig:fig1}
\end{figure}

In this work, we demonstrate a simple fact that making a cut (a vacuum region) in the bulk of a QAHI creates a single helical channel formed by two counter-propagating chiral edge modes. The QAHI with a cut is coupled to a superconductor as shown in Fig.~\ref{fig:fig1} to create MZMs at the ends of the cut. The width of the cut should be in the order of tens of nm which is comparable to the coherence length of a typical superconductor. In this geometry, the coupling of the edge modes is mediated by a gapped superconductor so that they will not hybridize too strongly to gap out each other. It is important to note that: 1) In this geometry, the superconducting gap does not have to overcome the bulk insulating gap of the QAHI to create MZMs and this greatly enlarges the topological regime. This is in sharp contrast to the two-dimensional QAHI/superconductor (QAHI/SC) heterostructure in which the bulk insulating gap must be smaller than the superconducting gap to create chiral Majorana modes on the edge or MZMs in the vortices \cite{Qi}.  2) The presence of the single helical channel is guaranteed by the bulk topological property of the QAHI and no other conducting channels are created. This can avoid the generation of low-energy in-gap bound states caused by multiple conducting channels as in the case of multi-channel nanowires \cite{Patrick2010}. 3) A large number of MZMs can be created by creating multiple cuts as demonstrated in Fig.~\ref{fig:fig4}. Topological qubits such as hexons can be fabricated \cite{Karzig2017}. 4) Braiding operations can be performed by tuning the coupling strength of the counter-propagating chiral edge modes. 


This paper is organized as follows. First, we show that the low energy physics of the two counter-propagating chiral edge modes coupled to a superconductor (as shown in Fig.~\ref{fig:fig1})  is described by a 1D topological superconductor. Then a lattice model for the QAHI/SC heterostructure with a cut is constructed. With this model, we explicitly demonstrate that MZMs are supported in the proposed setup.  Finally, we demonstrate how the MZMs can be detected by transport experiments and how to realize scalable topological quantum computation in this platform.

\emph{Topological superconductivity in QAHI/SC heterostructure with a narrow cut}.---
It is well-known that the chiral edge modes of two-dimensional QAHIs are protected by the topologically non-trivial bulk band structure \cite{Haldane1988, Nagaosa2003,Yurui2010}. When the width of the QAHI is reduced, the two counter-propagating chiral edge states residing on the opposite side of the QAHI will couple to each other and open a hybridization gap. However, if the separation of the two edge modes is not too short but only comparable to the localization length of the edge modes, an effective helical channel can be created inside the bulk gap ~\cite{Chuizhen}. Moreover, when superconductivity is induced on the helical channel by proximity to a superconductor, MZMs can be created in such a quasi-one-dimensional QAHI/SC heterostructure ~\cite{Chuizhen}. However, this scheme of creating MZMs requires the separation of the edge modes to be comparable to the coherence length of the superconductor~\cite{Chuizhen}. Experimentally, the localization length of the chiral edge modes in some QAHI can be much longer than the coherence length of the superconductor and the two edge modes can possibly couple to each other too strongly and gap out each other. Here we propose a simpler scheme to achieve a single helical channel by creating a narrow vacuum strip in the QAHI as depicted in Fig.~\ref{fig:fig1}.

As the chiral edge states are protected by the bulk topological property, the edge states will circumvent the narrow cut and generate a pair of counter-propagating edge modes as shown in Fig.~\ref{fig:fig1} (the green dashed arrows). The two chiral modes form a helical channel. The great advantage of this scheme is that: 1) the two chiral edge modes are in close proximity of each other so that a superconductor with relatively short coherence length can induce superconductivity on the edge modes. 2) The two edge states couple to each other only indirectly through the gapped superconductor so that they will not hybridize too strongly and gap out each other easily. 3) MZMs can be created whenever a pairing gap is induced on the helical channel and the pairing gap does not have to overcome the bulk insulating gap of the QAHI. Therefore, robust QAHIs with large bulk insulating gaps can be used to create Majorana fermions. As more and more QAHIs with large bulk gap are being proposed \cite{Binghai2014,Zhenyu2016,Chengxi2017,Yujun2020,Deng2020}, our scheme allows MZMs to be created on a robust topological platform. Therefore, our scheme is in sharp contrast to the two-dimensional QAHI/SC heterostructures which requires the pairing gap to overcome the bulk insulating gap which cannot be achieved in QAHIs with large insulating bulk gap \cite{Qi}.

\begin{figure*}
	\centering
	\includegraphics[width=1\linewidth]{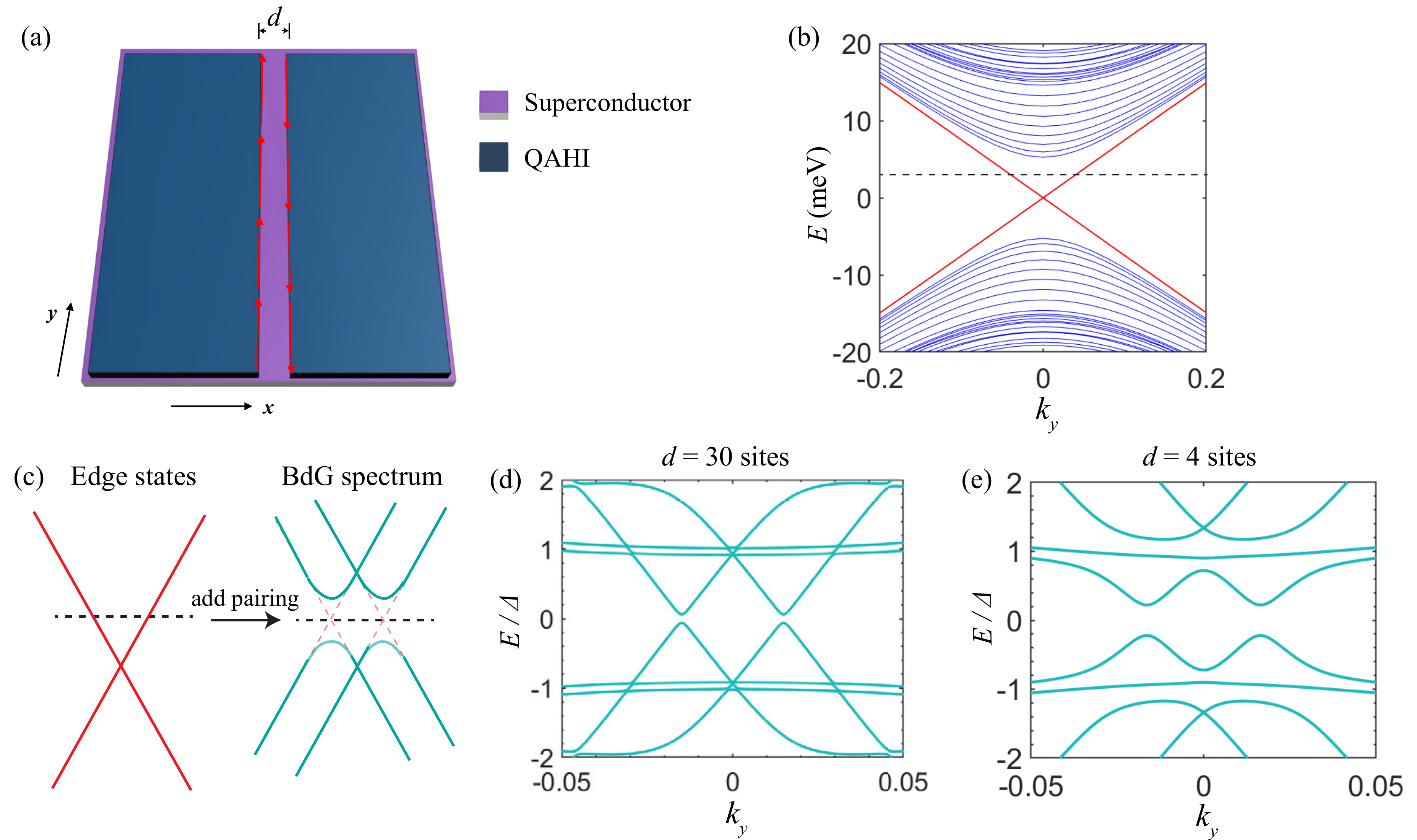}
	\caption{(a) A schematic plot of QAHI/SC geometry. A cut along the $y$-direction is created in the middle of QAHI. The red arrow denotes the edge states, $d$ denotes the separation between the two edge states.  (b) The energy spectrum for a QAHI with a cut, where the edge states near the cut are highlighted in red and the black dashed line indicates the position of chemical potential. Periodic boundary condition in the $y$-direction is imposed. (b) Schematic of the BdG energy spectrum when the superconducting pairing is introduced onto the edge states.  (d) and (e) show the energy $E$ (in unit of $\Delta$) as a function of $k_y$ for a sample with 160 lattice sites in the $x$-direction. In (d), width of the cut is 30 sites. In (e), the width of the cut is 4 sites.  In (b), (d), (e), the magnetization energy $M_z=10$ meV, the chemical potential  $\mu=3$ meV and the pairing potential $\Delta=1$ meV.}
	\label{fig:fig2}
\end{figure*}

To show that the helical edge modes in proximity to a superconductor exhibits topological superconductivity, we first build a low energy effective Hamiltonian for the two chiral edge modes near the cut region as shown in Fig.~\ref{fig:fig1}. The effective Hamiltonian can be written as $H_{eff}=\sum_{k_y}\Psi^{\dagger}_{k_y}\mathcal{H}_{BdG}(k_y)\Psi_{k_y}$ in the Nambu basis $\Psi_{k_y}=(\psi_{R,k_y},\psi_{L,k_y},\psi^{\dagger}_{R,-k_y},\psi^{\dagger}_{L,-k_y})^{T}$ and the Bogoliubov–de Gennes (BdG) Hamiltonian
\begin{equation}\label{Ham2}
\mathcal{H}_{BdG}(k_x)= v_0k_y\rho_z+c(k_y)\rho_x\xi_z-\mu\xi_z-\Delta_0\rho_y\xi_y,
\end{equation}
where $\psi_{R(L),k_y}$ is the annihilation operator of edge states with momentum $k_y$ at right (left) edge near the cut, $\rho$ and $\xi$ are Pauli matrices operating on (L,R) and particle-hole space respectively, $v_0$ denotes the velocity of chiral edge states, $c(k_y)=c_0+c_1k_y^2$ describes the coupling  between left-moving and right-moving edge states, $\mu$ denotes the chemical potential and $\Delta_0$ is the amplitude of proximitized superconducting pairing. Note that $\Delta_0$ depends on the coupling strength between two chiral edge states $c(k_y)$ implicitly. 

To see that Hamiltonian (\ref{Ham2}) describes a topological superconductor, we follow Ref.~\cite{Alicea2010,law_2020} to project out the pairing onto the conduction band $E_{+}=+\sqrt{v_{0}k_y^2+c^2(k_y)}-\mu$  and obtain an effective intra-band $p$-wave pairing 
 \begin{equation}
 \Delta_{eff}(k_y)=U^{\dagger}(k_y)\Delta_0i\rho_yU^{*}(-k_y)=\frac{v_0 k_y\Delta_0}{\sqrt{ ( v_0 k_y)^2+c(k_y)^2}},
  \end{equation}
 where the project operation $U(k_y)=\ket{E_{+}(k_y)}\bra{E_{+}(k_y)}$. This $p$-wave pairing results in a 1D topological superconductor \cite{Kitaev2001}. The topological region is $|c_0|<\sqrt{\Delta_0^2+\mu^2}$, which is determined by the gap-closing point of the BdG energy spectrum of $\mathcal{H}_{BdG}(k_y)$ at $k_y=0$ \cite{Kitaev1}. 
In the strong pairing limit, where $\Delta>c_0$, this system is always topological. In the weak pairing limit, where $\Delta_0$ is much smaller $|c_0|$,  the topological region is approximated as $|c_0|<\mu$. Intuitively, it means the system is topological as long as the chemical potential $\mu$ cuts through the edge states.

To demonstrate the validity of the effective Hamiltonian in Eq.~(\ref{Ham2}), we construct a tight-binding model for a QAHI/SC heterostructure with a cut and the geometry is illustrated in Fig.~\ref{fig:fig1}. As shown previously, the QAHI state can be realized by a thin film of magnetic doped topological insulator when the magnetic gap induced by the magnetic dopen is larger than the hybridization gap of the two Dirac surface states of the topological insulator \cite{Yurui2010}. Therefore, the QAHI Hamiltonian can be written as:
\begin{equation}
H_{QAHI}=\sum_{\bm{k}}\Phi_{\bm{k}}^{\dagger}( v_F k_y\sigma_x\tau_z- v_Fk_x\sigma_y\tau_z+m(\bm{k})\tau_x+M_z\sigma_z)\Phi_{\bm{k}}.
\end{equation}
Here, $\Phi^{\dagger}_{\bm{k}}=(\phi^{\dagger}_{t\bm{k}\uparrow},\phi^{\dagger}_{t\bm{k}\downarrow},\phi^{\dagger}_{b\bm{k}\uparrow},\phi^{\dagger}_{b\bm{k}\downarrow})$ is a four-component electron creation operator with momentum $\bm{k}$, where the subscripts $t\ (b)$ and $\uparrow(\downarrow)$ denotes the top (bottom) layer and spin up (down)  index respectively, $\sigma$ and $\tau$ are the Pauli matrices for spin and layer subspaces, $v_F$ denotes the Fermi velocity of topological insulator surface states, $m(\bm{k})=m_0+m_1(k_x^2+k_y^2)$ describes the hybridization between top and bottom surface states, and $M_z$ is the magnetization energy induced by the magnetic doping and external magnetic field. In the calculation, we set $ v_F=3$ eV$\cdot$\AA, $m_0=-5$ meV, $m_1=15$ eV$\cdot$\AA$^2$ \cite{Chuizhen}. The system is in the QAHI phase when the magnetization energy $M_z$ exceeds the hybridization energy $m_0$, where Chern number $\mathcal{C}=1$ for $M_z>|m_0|$ and $\mathcal{C}=-1$ for $M_z<-|m_0|$. In this phase, the system supports $|\mathcal{C}|$ chiral edge states. The tight-binding Hamiltonian for this QAHI is given in Appendix A. In the tight-binding model, we consider a strip of QAHI with periodic condition in the y-direction and 160-sites in the x-direction as shown in Fig.~\ref{fig:fig2}(a). Importantly, a cut along the y-direction is made in the middle of the QAHI to create a pair of counter propagating chiral edge modes. To eliminate the complication of the edge states on the outer sides of the sample, we utilize closed boundary condition on the left and right edges. The energy spectrum a QAHI with the geometry depicted in  Fig.~\ref{fig:fig2}(a) is shown in  Fig.~\ref{fig:fig2}(b). As expected, there are chiral edge modes inside the bulk gap. When the chemical potential is inside the bulk gap and superconductivity is induced, we expect the edge states will be gapped out as schematically shown in Fig.~\ref{fig:fig2}(c).

In our tight-binding model, we couple the top layer of QAHI to an extra layer of superconductor. The superconductor induce pairing as well as hybridization between the two edge modes. The energy spectrum of the QAHI/SC heterostructure with different edge state separations (denoted by $d$) are shown in Fig.~\ref{fig:fig2}(d) and Fig.~\ref{fig:fig2}(g). It is clear that when $d$ is much wider than the superconducting coherence length $\eta \approx v_F/\Delta_{0}$ of the superconductor which is estimated to be about 10 sites in the tight-binding model, the induced pairing gap of the system is small. However, when $d$ is comparable with $\eta$, the induced pairing gap is sizable. When a conventional s-wave pairing is induced on the two counter-propagating edge modes, the system immediately becomes a topological superconductor. In the next section, we demonstrate the presence of localized MZMs in the geometry of Fig.~\ref{fig:fig1}.

\begin{figure}
	\centering
	\includegraphics[width=0.8\linewidth]{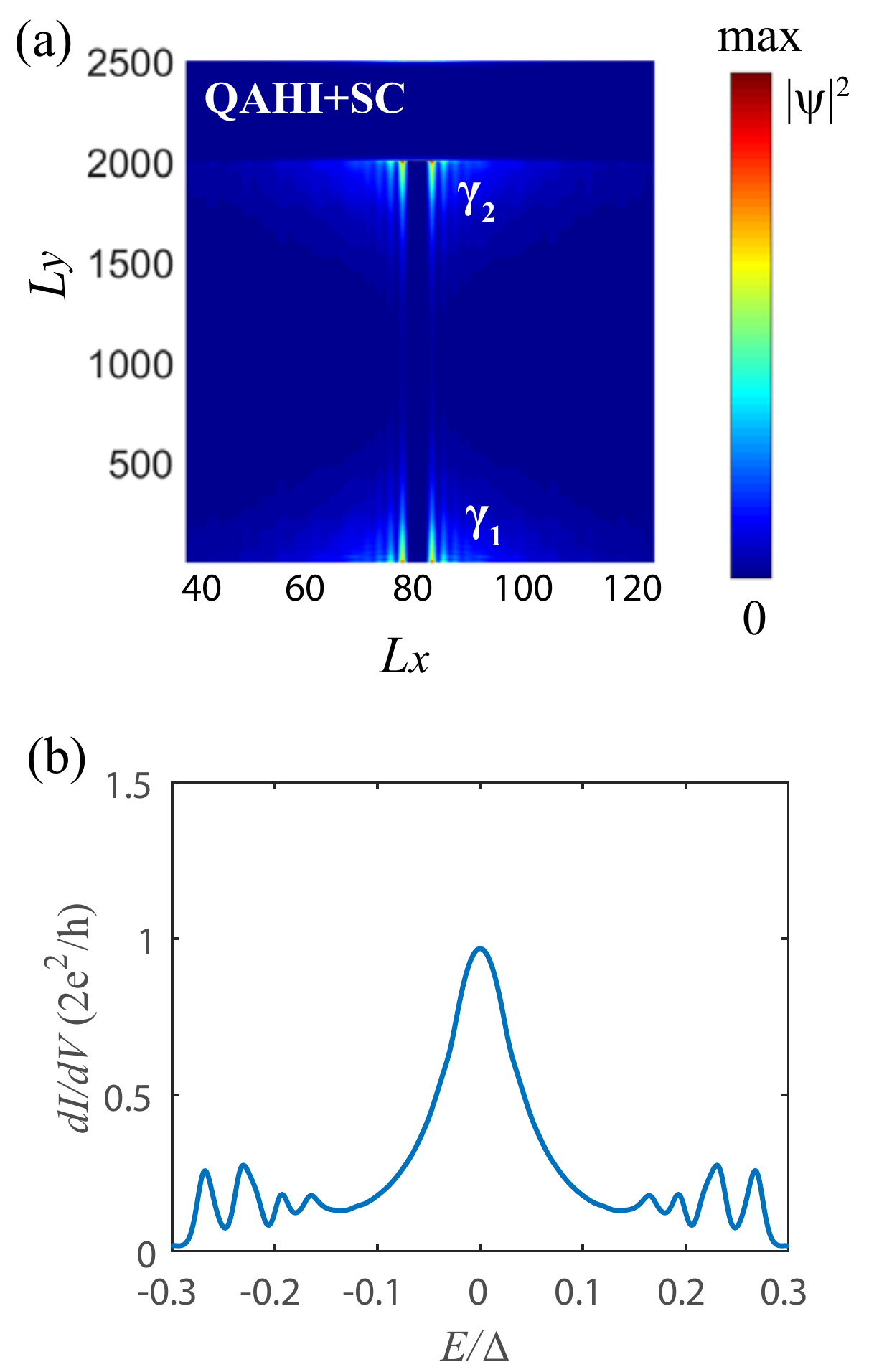}
	\caption{  (a) shows the localized wavefunction of MZMs near a cut, where the QAHI/SC sample is 2500 sites $\times$ 160 sites and a cut of 2000 sites $\times$ 4 sites is created in the middle with one end touching the edge. (b) The simulated tunneling spectra for the QAHI/SC geometry plotted in Fig.~\ref{fig:fig1}. A quantized $2e^2/h$ conductance is induced by the MZM. Here, the magnetization energy $M_z=10$ meV, the chemical potential  $\mu=3$ meV, and the temperature $T=0.05$ K are used. }
	\label{fig:fig3}
\end{figure}

\emph{Localized MZMs.}--- In this section, we explicitly demonstrate that MZMs are supported in the QAHI/SC heterostructure with a cut.  A tight-binding model with 2500 sites in the y-direction and 160 sites in the x-direction is constructed to describe the QAHI/SC heterostructure. A cut of 2000 sites long and 4 sites wide is made on the QAHI layer as depicted in Fig.~\ref{fig:fig1}. Due to the bulk topological property, the chiral edge mode will circumvent the vacuum region and propagate along the cut. This creates a pair of counter-propagating chiral modes locally on the two edges of the cut. When the QAHI is covered by a superconductor, the superconductor will mediate coupling as well as induce superconducting pairing on the chiral edge modes along the cut. This will result in a localized Majorana mode at the end of the cut as depicted in Fig.~\ref{fig:fig3}(a). 

It is important to note that, for the calculation with the tight-binding model, in order to eliminate the effect of the chiral edge modes which circulates at the boundary of the QAHI, we introduced periodic boundary conditions along both x and y directions such that the edge states appear along the vacuum strip only. In the realistic geometry of Fig.~\ref{fig:fig1}, there are gapless edge modes entering and leaving the vacuum strip region. Therefore, Majorana mode $\gamma_1$ will be delocalized and the wavefunction will merge with the wavefunction of the gapless chiral edge modes. On the other hand, Majorana mode $\gamma_2$ will remain localized which can be detected by STM measurements. The details of the tight-binding model is given in the Appendix.


\begin{figure*}
	\centering
	\includegraphics[width=1\linewidth]{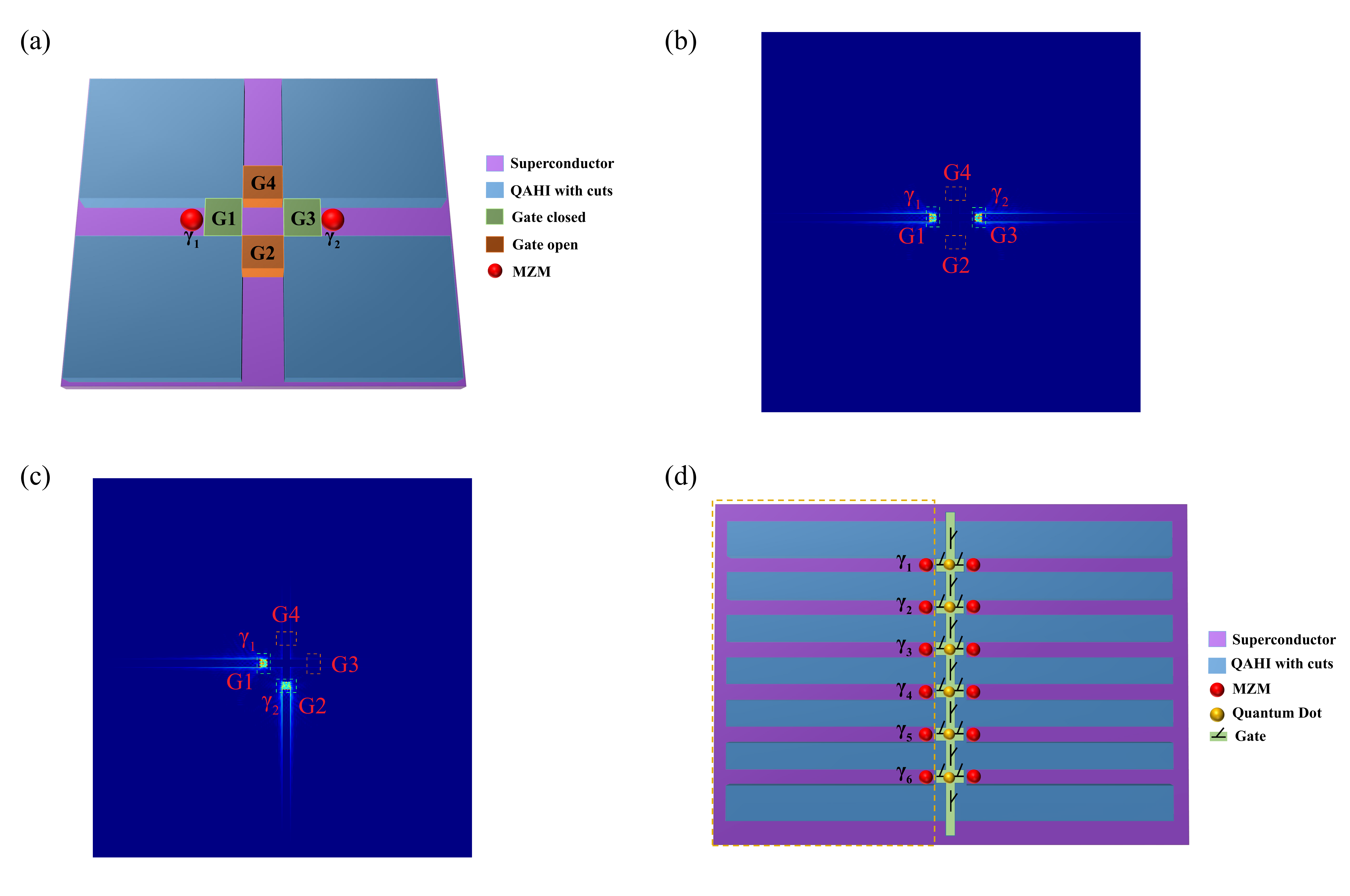}
	\caption{(a) A cross-shape cut on a QAHI/SC heterostructure through which braiding operations of MZMs $\gamma_1$ and $\gamma_2$ can be performed. Gn (n from 1 to 4) label the four gates made from semiconducting materials. (b) and (c) show the wavefunctions $|\psi|^2$ of MZMs localized at the closed gates with the geometry displayed in (a). G1 and G3 are closed and other gates are open in (b), while G1 and G2 are closed and other gates are open in (c). (d) A schematic picture of a hexon based on the QAHI/SC heterostructure.} 
	\label{fig:fig4}
\end{figure*}

\emph{Experimental detection.}---Tunneling experiments can be conducted to detect the MZMs. We propose a simple geometry to detect the MZMs in this setup, as shown in Fig.~\ref{fig:fig1}. A cut is made perpendicularly on the one edge of the QAHI sample and the superconductor covers the vacuum region to create MZMs.  A macroscopic lead is attached to opposite edge of the QAHI and the superconductor is grounded. An incoming electron mode from the lead comes into the QAHI is reflected to a hole mode by the MZM, which gives rise to a resonant Andreev reflection \cite{law,Wimmer_2011}. We numerically simulated the tunneling differential conductance of the system with the recursive Green's function methods \cite{Patrick1981,Sun_2009,liujie2012,liujie2013}. The differential conductance as a function of voltage difference between the lead and the grounded superconductor is shown in Fig.~\ref{fig:fig3}(b). It clearly shows a quantized zero bias peak with value $2e^2/h$ which is expected for a topological superconductor. The localized MZMs can also be detected by scanning tunneling spectroscopy experiments.

 \emph{Towards scalable topological quantum computation.}--- After discussing how MZMs can be created, in this section, we propose two ways to manipulate the MZMs for topological quantum computation. 	In the superconductor-coupled semiconducting nanowire case, the braiding of MZMs can be achieved in a T-junction or a cross-shape junction \cite{Alicea2011}. However, such T-junction or cross-shape junctions are difficult to fabricate experimentally, especially when a network of such junctions is needed for practical quantum computation \cite{Karzig2017,Alicea_2016}.   Here, we point out that junctions with complicated geometries can be replaced by cuts on a QAHI/SC heterostructure.  As an illustration, we schematically show a cross-shape cut in Fig.~\ref{fig:fig4}(a), where there are four gates near the center of the cross. The area under the four gates can be made from semiconducting materials such that by tuning the chemical potential, the semiconducting material can be conducting or insulating. When the gates are open, the area under the gate is insulating. When the gates are closed, the area under the gate is metallic and mediate strong coupling between the two edge modes.

As shown in Fig.~\ref{fig:fig4}(b), Gate 1 and Gate 3 are closed so that the two chiral edge modes hybridize strongly near Gate 1 and Gate 3. This creates a topologically trivial region near the two gates. As a result, two localized MZMs are created near the center of the junction. Moreover, when Gate 1 and Gate 2 are closed while Gate 3 and Gate 4 are open, the MZM $\gamma_2$ is moved from one arm of the cross to another arm as shown in Fig.~\ref{fig:fig4}(c). Appropriate opening and closing of the gates will lead to non-Abelian braiding of the MZMs as in Ref.~\cite{Chuizhen}. The details of the parameters for the simulation can be found in the Appendix.
	
 Recently, it was proposed that the braiding of MZMs could be replaced by the parity measurement of MZMs. This scheme is called the measurement-based topological quantum computation \cite{Bonderson2008,Fuliang2015,Fuliang2016}. The hexon and tetron architectures proposed in Ref.~\cite{Karzig2017} to achieve the measurement-based topological quantum computation were initially based on semiconducting nanowires. Here, we suggest that the QAHI/SC heterostructure is an experimentally feasible platform for the proposed architectures. The so-called \lq\lq{}one-sided\rq\rq{} hexon built on QAHI/SC is illustrated in Fig.~\ref{fig:fig4}(d). The six MZMs ($\gamma_n$ where $n$ goes from 1 to 6) of the hexon are localized at the ends of the cuts of the QAHI.  It is important to note that the MZMs located in the middle of the QAHI are separated from the low energy modes residing on the circumference of the QAHI by fully gapped QAHI regions or fully gapped superconducting regions so that the gapless edge modes will not affect the fermion parity of the hexon formed by the six localized MZMs. As proposed in Ref.~\cite{Karzig2017}, the full set of the single-qubit Clifford gates can be generated by sequential measurements of fermion parities of pairs of MZMs. 

\emph{Conclusions}---
	In this work, we  propose a simple platform for realizing MZMs using QAHI/SC heterostructures with cuts. In this platform, the single helical channel required for creating localized MZMs is protected by the bulk topological property of the QAHI. The created MZMs can be detected by transport experiments and braiding operations can be performed by gating. Complicated qubits which require the presence of multiple MZMs can be realized. 
	
\emph{Acknowledgments.}--- KTL acknowledges the support of the Croucher Foundation, the Dr. Tai-chin Lo Foundation and the HKRGC through grants C6025-19G, C6026-16W, 16310219  and 16309718.

\begin{appendix}
%

	\section{Tight-binding model of a QAHI/SC heterostructure  with a cut}	
	In this section, we present the tight-binding model that is used in main text to simulate a QAHI/SC heterostructure with a cut. We consider a three-layer geometry, where the first two layers are for the QAHI and the third layer is for the parent superconductor. 
	The lattice model of the QAHI is given by
	\begin{eqnarray}
	H_{QAHI}=\sum_{k_y,j}&&\Phi_{j,k_y}^{\dagger}[v_F\sin k_y\sigma_x\tau_z+M_z\sigma_z+(m_0+4m_1\nonumber\\
	&&-2m_1\cos k_y)\tau_x]\Phi_{k_y,j}+\Phi_{k_y,j}^{\dagger}(iv_F/2\sigma_y\tau_z\nonumber\\
	&&+m_1\tau_x)\Phi_{k_y,j+1}+\text{H.c.},
	\end{eqnarray}
	where   $\Phi^{\dagger}_{j,k_y}=(\phi^{\dagger}_{tk_y\uparrow,j},\phi^{\dagger}_{tk_y\downarrow,j},\phi^{\dagger}_{bk_y\uparrow,j},\phi^{\dagger}_{bk_y\downarrow,j})$ and $\phi^{\dagger}_{t(b)k_ys,j}$ creates an electron at site $j$ of top (bottom) layer with momentum $k_y$, spin $s$.  We simulate the cut on the QAHI layer by a stripe of vacuum area with very high on-site potential and zero coupling to other sites.
	In the calculation, for the sake of convenience, the lattice constant $a$ is taken as 4 nm \cite{Chuizhen} and other parameters can be transformed in units of energy accordingly, such as $v_F\rightarrow v_F/a$, $m_1\rightarrow m_1/a^2$ \textit{etc.}.
	
	The lattice model of the parent superconductor layer is given by
	\begin{eqnarray}
	H_{SC}=\sum_{k_y,s,1\leq j\leq N} &&c_{k_y,j,s}^{\dagger}(4t_{sc}-2t_{sc}\cos k_y-\mu_{sc})c_{k_y,j,s}\nonumber\\
	&&-t_{sc}c_{k_y,j,s}^{\dagger}c_{k_y,j+1,s}+\text{H.c.}\nonumber \\ 
	&&+\Delta c^{\dagger}_{k_y,j,s}(i\sigma_y)_{ss'}c^\dagger_{-k_y,j,s'}+\text{H.c.},
	\end{eqnarray}
	where $t_{sc}$, $\mu_{sc}$ respectively denote the hopping amplitude and the chemical potential of the superconductor layer, $\Delta$ is the superconducting pairing potential. In the calculation, we set $t_{sc}=v_F/a$, $\mu_{sc}=t_{sc}$, $\Delta=1$ meV.
	
	The coupling between the QAHI and the superconductor is given by
	\begin{equation}
	H_c=\sum_{k_y,s,j}\Gamma_c\phi^{\dagger}_{bk_y,j,s}c_{k_yj,s}+\text{H.c.}
	\end{equation}
	Here $\Gamma_c$ denotes the coupling strength and only the coupling between the bottom layer of QAHI and the superconductor layer is considered. In the calculation, we set $\Gamma_c=v_F/4a$.
	
	The total Hamiltonian describing the QAHI/SC heterostructure with a cut is written as
	\begin{equation}
	H_{t}=H_{QAHI}+H_{SC}+H_{c}+H_{\mu}\label{Ham3},
	\end{equation} 
	where an ancillary part $H_\mu$ is added to tune the chemical potential of the whole system, which is given by
	\begin{equation}
	H_{\mu}= -\sum_{k_y, j}\mu(\Phi_{k_y,j}^{\dagger}\Phi_{k_y,j} + c_{k_y,j}^{\dagger}c_{k_y,j}).
	\end{equation}

	In Fig.~\ref{fig:fig4}(b) and Fig.~\ref{fig:fig4}(c), the gates are simulated by an ultra-simplified semiconductor with the hopping parameter $t_{gate}=v_F$ and the gap size $m_{gate}=20$ meV. In the calculation of these subfigures, to make the MZMs more localized, we have adopted a smaller $v_F$ which is one third of the original $v_F$ mentioned in the main text. The sample size is 500-by-500 lattice sites, and periodic boundary conditions on both $x$- and $y$-directions are adopted.

\end{appendix}
	\bibliographystyle{apsrev4-1} 
	\bibliography{Reference}

\end{document}